\documentclass[a4paper,11pt]{article}
\usepackage[textwidth=15cm]{geometry}
\usepackage{graphicx,color}  
\usepackage{amsfonts,amssymb,amsthm,amsmath}
\usepackage{subcaption}
\usepackage[colorlinks,citecolor=red,urlcolor=blue,bookmarks=false,hypertexnames=true]{hyperref}

\newtheorem{example}{Example}[section]

\begin{document}
\title{
  Effective nonlinear Ehrenfest hybrid quantum-classical dynamics
}

\author{J. L. Alonso$^{1,2,3}$, C. Bouthelier-Madre $^{1,2,3}$, J. Clemente-Gallardo $^{1,2,3}$, \\ D. Martínez-Crespo $^{1,3}$ and J. Pomar$^1$}
\date{$^1$ Departamento de F\'{\i}sica Te\'orica, Universidad de Zaragoza,  Campus San Francisco, 50009 Zaragoza (Spain) \\
$^2$ Instituto de Biocomputaci{\'{o}}n y F{\'{\i}}sica de Sistemas Complejos (BIFI), Universidad de Zaragoza,  Edificio I+D, Mariano Esquillor s/n, 50018 Zaragoza (Spain) \\
$^3$ Centro de Astropart{\'{\i}}culas y F{\'{\i}}sica de Altas Energías (CAPA), Departamento de F\'{\i}sica Te\'orica, Universidad de Zaragoza,  Campus San Francisco, 50009 Zaragoza (Spain)
}
\maketitle


\abstract{The definition of a consistent evolution equation for statistical hybrid quantum-classical systems is still an open problem.  In this paper we analyze the case of Ehrenfest dynamics on systems defined by a probability density and identify the relations of the non-linearity of the dynamics with the obstructions to define a consistent dynamics for the first quantum moment of the distribution. This first quantum moment represents the physical states as a family of classically-parametrized density matrices $\hat \rho(\xi)$, for $\xi$ a classical point; and it is the most common representation of hybrid systems in the literature. Due to this obstruction, we consider higher order quantum moments, and argue that only a finite number of them are physically measurable. Because of this, we propose an effective solution for the hybrid dynamics problem based on approximating the distribution by those moments and representing the states by them.}

\maketitle

\section{Introduction} 
\label{Intro}

It is an accepted fact that the most accurate physical theories describing Nature are quantum. Nonetheless, it is also well known that {\color{black} ab-initio full quantum theories are not useful from the practical point of view, because of their complexity}. One of the choices is then to approximate the full quantum model by a simpler one where as many degrees of freedom as possible are modelled as classical systems. The paradigmatic example of this situation is the model of a molecule. Based on the application we have two options: either to consider all the degrees of freedom as classical (this is the case in most frequent molecular dynamics simulation methods), or to {\color{black} create a hybrid quantum-classical model where only the valence electrons, responsible of the chemical properties of the molecule, are modelled as quantum objects, the rest of degrees of freedom being represented by classical variables. In this second case, we define a more accurate description of the chemical properties, with a  much simpler model than the full-quantum one}. Nonetheless, the definition of an accurate  hybrid dynamical system is not an easy task.  There exist several approaches to define hybrid dynamical models for molecular systems (see, for instance, \cite{Yonehara2012} for a recent review). We can classify them based on different properties, but, attending only to the definition of the dynamics, we may consider, among others:
\begin{itemize}
	\item  those based on hybrid dynamics on the space of hybrid states (\cite{zhuNonBornOppenheimerLiouvillevonNeumann2005,Jasper2005,Jasper2006,alonsoStatisticsNoseFormalism2011,alonsoEhrenfestDynamicsPurity2012,alonsoEhrenfestStatisticalDynamics2018,Agostini2014}), 
	\item those which are algorithmic (see \cite{tullyMolecularDynamicsElectronic1990,Tully1998})
	\item  or others obtained as suitable limit equations of the full-quantum dynamics (\cite{Prezhdo1997,Kapral1999,Kapral2015,nielsenNonadiabaticDynamicsMixed2000,nielsenStatisticalMechanicsQuantumclassical2001}. 
\end{itemize}

If we enlarge our scope and consider other hybrid dynamical models, we can also find those considering the problem of measurements of quantum systems with classical devices \cite{sherryInteractionClassicalQuantum1979,Buric2012} and  other type of systems (see \cite{Hall2008,Peres2001,Elze2013,Diosi2014} and references therein).
  
One of the most common choices is the Ehrenfest dynamics, where it is simple to track the accuracy of the hybrid model with respect to the original full quantum one within a semiclassical description \cite{Bornemann1996}. Ehrenfest equations represent a dynamical system defined on the cartesian product of the classical and quantum phase spaces $M_C\times M_Q$. This hybrid phase space  contains the hybrid pure states, i.e., those states where the classical and the quantum degrees of freedom are completely determined. Our group proved \cite{alonsoStatisticsNoseFormalism2011} that this system admits a Hamiltonian description  with a suitable hybrid Poisson bracket and a hybrid Hamiltonian function, which combines the classical and the quantum energies. Nonetheless, molecular systems do not usually have completely determined dynamical states, since initial conditions are impossible to fix. Therefore, a statistical description appears to be the most reasonable one. In \cite{alonsoStatisticsNoseFormalism2011} we also used the Hamiltonian structure of Ehrenfest equations to define a  statistical model having Ehrenfest equations as dynamics for the microstates. In this statistical description, the state of the system is defined as a probability distribution on the hybrid phase-space following a Liouville equation.  This is a consistent mechanism to define a statistical mechanical system associated to a well defined thermodynamics (see \cite{balescuStatisticalDyamicsMatter1997}).

In the last years, several relevant applications of this scheme have been published (\cite{alonsoEhrenfestDynamicsPurity2012,alonsoEhrenfestStatisticalDynamics2018}) but also some limitations have appeared. Among them, the most relevant one may be the difficulty to write an entropy function and the corresponding notion of canonical ensemble \cite{alonsoEntropyCanonicalEnsemble2020} or, in more general terms, of an equilibrium thermodynamics with a well defined temperature. 

Our goal in this paper is a careful analysis of the origin of these difficulties, and a discussion of possible solutions and their implications.  From the mathematical point of view, the problems are associated with the incompatibility of the notion of hybrid state as a probability density on the phase space and the definition of hybrid entropy, which require an alternative notion of state. Instead of the probability density described above, entropy can be formulated by considering its first quantum moment, which defines the type of hybrid state which has been used extensively since the early eighties (see \cite{aleksandrovStatisticalDynamicsSystem1981}) and is used in many of the references presented above. But the challenge is to define a consistent master equation for this object, since the Liouville equation of the probability density does not preserve the first moment of the density.  We will see this property in detail, and define a series of equations for the different moments which are equivalent to the  Liouville equation for the probability distribution. Nonetheless, depending on the physical situation, not all the moments are necessary to define the physically relevant averages and their evolutions. For those situations where only a finite number of them are physically relevant, just a finite system of equations characterize the evolution of the quantum moments and an effective nonlinear master equation can be defined. This is the main result of the paper. 

It is important to remark that the properties of the Hamiltonian structure of the microstate dynamics, such as having two independent symplectic or Poisson structures (as Ehrenfest dynamics), is the true origin of these properties. The particular Hamiltonian function, and hence the detailed properties of the dynamics are not that relevant. Hence, most of the conclusions of our paper shall be true in any Hamiltonian system sharing independent structures for the classical and quantum subsystems.

The scheme of the paper is as follows. First, in Section \ref{sec:Ehrenfest_dynamics} we will summarize the construction of the Hamiltonian structure for Ehrenfest equations and its statistical extension. We will discuss that its Hamiltonian nature implies that only a finite number of quantum moments of the hybrid distribution are physically measurable.  Then, Section \ref{sec:Hybrid_states} analyzes the properties of the different definitions of the hybrid statistical state and justifies why Liouville equation can not be restricted to any of the quantum moments.  The reason for that is the nonlinearity introduced in Ehrenfest equations by the classical degrees of freedom. Nonetheless, we can encode the full Liouville equation in a series of coupled differential equations of the full set of moments. But since only a finite number of moments are physically meaningful, we can approximate the behavior of Liouville equation by a finite number of those differential equations, suitably adapted. This is the main result of the paper.  Finally, Section \ref{sec:Conclusions} summarizes the main contributions of the paper and discusses future lines of research based on them.

\section{Ehrenfest dynamics: from pure states to distributions}
\label{sec:Ehrenfest_dynamics}

\subsection{Ehrenfest equations on pure states}
Let us consider first the problem of Ehrenfest dynamics for pure states. We assume then that the state of the hybrid system is characterized by a pair $(\xi, \rho_\psi)$, where $\xi=(\vec q, \vec p)\in M_C$ is a point on the classical phase space and $\rho_\psi$ specifies the pure state of a quantum system by a one-dimensional projector determining a point in the complex projective space  $\mathcal{PH}$ of a certain Hilbert space $\mathcal{H}$. 
Ehrenfest equations define a set of coupled differential equations describing the evolution of these degrees of freedom as 
\begin{eqnarray}
\label{eq:ehrenfest_equations}
\dot q^k=& \frac {p^k} M  \nonumber \\ 
\dot p^k=& -\frac{\partial \mathrm{Tr} (\rho_\psi \hat H(\xi))} {\partial q^k} \\ 
i\dot \rho_\psi=&[\hat H(\xi), \rho_\psi] \nonumber
\end{eqnarray}
where $\hat H(\xi)$ represents the quantum energy as a self-adjoint family of operators on $\mathcal{H}$ parametrized by the classical degrees of freedom {\color{black} written in Darboux coordinates $q^k, p_j$ for $j,k=1,\cdots n$ for the $2n$--dimensional symplectic manifold $M_C$ (see \cite{abrahamFoundationsMechanics1978} for the definition of these concepts)}. The simplest image for this operator is the electronic Hamiltonian for the valence electrons of a molecular system which depends on the positions of the charges of the protons and inner electrons of the atoms which are modelled by the classical degrees of freedom. In \cite{Bornemann1996} it is proved that this system of equations defines a dynamical system whose solutions approximate the solutions of the Schrödinger equation for a full-quantum molecular Hamiltonian, the accuracy depending on the ratio of the masses of the quantum and the classical particles and the ratio of the width of the nuclear wave-packet with respect to the natural length of the problem.

The classical phase space $M_C$ is assumed to be endowed with a symplectic form $\omega_C$. Hence, its functions $C^\infty(M_C)$ can be endowed with a canonical Poisson bracket $\{\cdot, \cdot\}_C$.  On the other hand, the quantum degrees of freedom correspond to the points of the set of pure density matrices of a quantum system (which we will denote in the following as $M_Q$), which being a Kähler manifold, is also endowed with a canonical Poisson structure $\{ \cdot, \cdot \}_Q$ (for details see \cite{carinenaGeometrizationQuantumMechanics2007,clemente-gallardoSpaceDensityStates2007,carinenaTensorialDynamicsSpace2017}). For the sake of simplicity we will consider that the quantum system is finite dimensional, and hence that $M_Q$ is a finite dimensional Kähler manifold. From the point of view of applications, this is not a serious constraint, since numerical simulation, which is usually the final goal of the hybrid model, requires a finite dimensional system. 

In conclusion, we can endow the set of functions of the hybrid phase space $M_C\times M_Q$ with a Poisson structure by combining the classical and the quantum brackets (for simplicity, we will fix $\hbar=1$ in the following) obtained from the symplectic forms on $M_C$ and $M_Q$:
\begin{equation}
\label{eq:hybrid_poisson}
\{\cdot , \cdot\}_H:=\{ \cdot, \cdot \}_C+\{ \cdot, \cdot \}_Q.
\end{equation}

From the results in \cite{alonsoStatisticsNoseFormalism2011}, it is immediate to prove that Ehrenfest equations can be given a Hamiltonian description with respect to the hybrid Poisson bracket $\{ \cdot, \cdot\}_H$ and the Hamiltonian function
{\color{black}
  \begin{equation} 
\label{eq:ehrenfest_hamiltonian}
f_{\hat H(\xi)}:= \mathrm{Tr} \left (\rho_\psi (H_C(\xi)\hat {\mathbb{I}}+\hat H(\xi) )\right ), 
\end{equation}
}where we represent as $H_C(\xi)$ the general expression of the energy of the classical degrees of freedom (in Equations \eqref{eq:ehrenfest_equations} it would reduce to just the classical kinetic term but more general situations with arbitrary classical potentials can also be considered); and $\hat {\mathbb{I}}$ represents the identity operator. 

Thus, the integral curves of the Hamiltonian vector field 
{\color{black} 
\begin{equation} 
\label{eq:Hamiltonian_vf}
X_{f_{H}}:= \{\cdot, f_H \}, 
\end{equation}
}correspond to the solutions of Ehrenfest equations. The dynamics preserves the purity of the quantum states but it is non-linear in general, because of the classical degrees of freedom (the term $H_C(\xi)$ in Equation \eqref{eq:ehrenfest_hamiltonian}). For more details see \cite{alonsoStatisticsNoseFormalism2011, alonsoEhrenfestDynamicsPurity2012}.

\subsection{Ehrenfest dynamics for statistical systems}


\subsubsection{The general case}
From the definition of the hybrid Hamiltonian, we can consider the general notion of hybrid observable, representing the physical magnitudes written in terms of the classical and quantum degrees of freedom.
Generally speaking, we can think at the hybrid observables as the tensor product of the $C^*$--algebra of classical functions and the $C^*$--algebra of quantum operators. For technical reasons, we will ask the classical functions to have compact support for integrals to be well defined. In practical applications, most frequently computer simulations, this does not seem to be a strong requirement. If we use the geometrical formulation described in \cite{alonsoStatisticsNoseFormalism2011}, we can consider that the space of hybrid observables can be represented by the functions on $M_C\times M_Q$ which correspond to a pair of classical and quantum observables, i.e., functions of the form
{\color{black}
\begin{equation}
\label{eq:hybrid_function}
f_{\hat A(\xi)}:=\mathrm{Tr} \left ( \hat \rho_\psi \hat A(\xi) \right ).
\end{equation}
}
In \cite{alonsoStatisticsNoseFormalism2011} we proved that it is possible to define a statistical mechanical formalism for hybrid systems by introducing a measure in this set of observables. If we take a reference measure $d\mu_{QC}$ on the hybrid phase space (for instance, the sympletic volume), the chosen measure can be written as a probability density function $F_{QC}(\xi, \rho_\psi)$ with respect to it, defining the average of a certain magnitude $\hat A(\xi)$ as
{\color{black}
  \begin{equation}
\label{eq:average}
\langle \hat A(\xi)\rangle:=\int_{M_C\times M_Q} d\mu_{QC} (\xi, \rho_\psi) F_{QC} (\xi, \rho_\psi) f_{\hat A (\xi)}.
\end{equation}
}Naturally, the measure must be well defined and therefore the density must satisfy that 
\begin{equation}
\label{eq:conditions_density}
F_{QC} \geq 0; \qquad \int_{M_{C}\times M_{Q} } d\mu_{QC} (\xi, \rho_\psi) F_{QC}=1. 
\end{equation}

\subsubsection{The first quantum moment}
Notice, though, that in physical terms, only the first quantum moment of $F_{QC}$ is necessary to determine the average value of a hybrid observable of the form of Equation \eqref{eq:hybrid_function}. Indeed, the first quantum moment of $F_{QC}$ determines a family of quantum operators indexed by the classical variables $\xi$ in the form:
\begin{equation}
\label{eq:rho_xi}
\hat \rho (\xi)= \int_{M_{Q}} d\mu_Q  (\rho_\psi)F_{QC}(\xi, \rho_\psi) \rho_\psi,
\end{equation}
where $d\mu_Q$ is a measure on $M_Q$ (for instance, the quantum symplectic volume).
Average values of hybrid observables can then be computed as 
\begin{equation}
\label{eq:average_rho}
\langle \hat A(\xi)\rangle =\int_{M_C} d\mu_C (\xi) \mathrm{Tr} \left ( \hat \rho(\xi)  \hat A(\xi) \right ),
\end{equation}
where $d\mu_C$ is a measure on $M_C$ (for instance, the classical symplectic volume) and then $d\mu_{QC}=d\mu_Q d\mu_C$. {\color{black} Notice that we use the term \textit{quantum moment} since, considering a basis for states $\rho_\psi$, the coordinates of $\hat \rho(\xi)$ correspond, precisely, to the moments of the probability distribution $F_{QC}$ considered as a function of the coordinates of the quantum projectors.}

Furthermore, from the mathematical point of view, we can relate this first quantum moment with the marginal and conditional probabilities of the original density $F_{QC}$. As a bivariate hybrid measure, we can consider its classical marginal probability and the corresponding quantum conditional one:
\begin{itemize}
	\item The classical marginal probability $F_C(\xi)$ can be obtained as 
	\begin{equation}
	\label{eq:marginal}
	F_C(\xi)=\mathrm{Tr} \hat \rho(\xi)=\int_{M_Q} d\mu_Q (\rho_\psi) F_{QC}(\xi, \rho_\psi).
	\end{equation}
	\item The quantum conditional probability for a classical value $\xi\in M_C$ is a purely quantum distribution. {\color{black} We can consider it to be a distribution over the quantum phase space with some density $F^{cond}_\xi(\rho_\psi)$ on the quantum phase space $M_Q$.   }
  As such, Gleason theorem \cite{gleasonMeasuresClosedSubspaces1957} ensures that there exists a well defined quantum density matrix $\hat \rho_\xi$ which is able to determine the average value of any quantum observable (for a fixed value of the classical variable $\xi$).
\end{itemize}
{\color{black} As it was explained in \cite{alonsoEntropyCanonicalEnsemble2020}, 
s}tandard probability theory allows us to write the operator $\hat \rho(\xi)$ as the product of these two objects, i.e., 
\begin{equation}
\label{eq:factorization_rhoxi}
\hat \rho(\xi)=F_C(\xi) \hat \rho_\xi.  
\end{equation}

Notice that the resulting object is not normalized as a density matrix, since 
\begin{equation}
\label{eq:trace_rho_xi}
\mathrm{Tr} \hat \rho(\xi)=F_C(\xi); \qquad \int_{M_C}d\mu_C (\xi) \mathrm{Tr} \hat \rho(\xi)=1.
\end{equation}

This operator $\hat \rho(\xi)$ is the usual representation of a hybrid state in the literature, since the early eighties \cite{aleksandrovStatisticalDynamicsSystem1981}; and it is the usual choice in most of the references we presented in the Introduction. Notice, though, that from the probabilistic point of view, it just {\color{black} represents an element of the dual space to the set of}  hybrid observables of the form of Equation \eqref{eq:hybrid_function}. If we consider the $C^*$--algebra defined by the operators of that type, operator $\hat \rho(\xi)$ defines a state for such an algebra.  From this point of view, this {\color{black} first quantum moment  of  $F_{QC}$  is the only relevant part of the probability distribution if we restrict the physical magnitudes to functions of the form \eqref{eq:hybrid_function}, since higher moments do not contribute to any average value}. Hence we can define an equivalence relation in the space of hybrid measures by the first moment: two hybrid measures represent the same physical state if and only if their first quantum moments coincide. 

Furthermore, this first moment $\hat \rho(\xi)$ is also able to capture the mutual exclusivity of hybrid events (see \cite{alonsoEntropyCanonicalEnsemble2020}). Indeed, notice that the points of the hybrid phase-space $M_C\times M_Q$ do not represent mutually exclusive events from the probabilistic point of view, as it happens with a classical manifold. While a classical system at a point $\xi_1$ can not be at the same {\color{black} time} at point $\xi_2$ (the probabilities of one case and the other are independent hence), it is not incompatible for a quantum system to be at point $\rho_{\psi_1}$ and at point $\rho_{\psi_2}$ unless $\langle \psi_1, \psi_2\rangle=0$ (probabilities are not independent then).  From that point of view, representing a {\color{black} hybrid} state with the density $F_{QC}$ makes very difficult to define a {\color{black} hybrid} entropy function ({\color{black} which requires of a correct representation of independent events which, as we just argued, is not achieved on $M_C\times MQ$). The first moment representation $\hat \rho(\xi)$, on the other hand, represents well the independence of hybrid events (as von Neumann entropy does for quantum systems)} and allows us to define a well-behaved entropy function. From it, we can also use the MaxEnt principle to define a consistent notion of canonical ensemble and therefore a mechanism to make finite-temperature numerical simulations of hybrid systems  (see \cite{alonsoEntropyCanonicalEnsemble2020, Alonso2021} for details).

\subsubsection{Considering dynamics: the necessity of  higher order moments}
From a static point of view, the picture above is perfect. For the natural set of hybrid observables, we can consider a state corresponding to a probability density $F_{QC}$ on the hybrid phase space $M_C\times M_Q$, but only the first quantum moment is necessary to define averages of physical magnitudes. That first moment also allows us to define a consistent notion of hybrid entropy. The problem arises when we want to consider a dynamical framework for statistical averages or even to equilibritum thermodynamics. 

In this paper, we are going to consider Ehrenfest dynamics as the candidate for the dynamics of the statistical microstates.  It is immediate to verify that dynamical equation \eqref{eq:ehrenfest_equations} defines also a dynamics on the set of hybrid magnitudes, which can be written by means of the Hamiltonian vector field $X_{f_{H}}$. Indeed, we can write:
{\color{black}\begin{equation}
\label{eq:hybrid_heisenberg}
\frac{d}{dt} f_{\hat A(\xi)}(t)= X_{f_{\hat H(\xi)}} f_{\hat A(\xi)}:=\{ f_{\hat A(\xi)}, f_{\hat H(\xi)} \}_H. 
\end{equation}
Clearly, a function of the form $f_A(\xi, \rho_\psi;t):=f_A(\xi(t), \rho_\psi(t))$ where $(\xi(t), \rho_\psi(t))$ is a solution of Ehrenfest equations,  is itself  a solution of Equation \eqref{eq:hybrid_heisenberg} with initial condition $f_A(\xi, \rho_\psi;0)=f_A(\xi, \rho_\psi )$.
Therefore, the system has solutions, and we can consider the time-dependence of statistical averages 
\begin{equation}
	\label{eq:average-t}
	\langle \hat A(\xi)\rangle (t):=\int_{M_C\times M_Q} d\mu_{QC}(\xi, \rho_\psi) F_{QC} (\xi, \rho_\psi) f_{\hat A(\xi)} (t),
	\end{equation}
}and define out-of-equilibrium statistical mechanics of hybrid systems.

{\color{black}
But it is important to realize that  hybrid observables of the form \eqref{eq:hybrid_function} do not define a Poisson subalgebra for the hybrid Poisson bracket \eqref{eq:hybrid_poisson}. Indeed, given $f_{\hat A(\xi)}$ and $f_{\hat B(\xi)}$ defined as Equation \eqref{eq:hybrid_function}, in general there exists no function $f_{\hat D(\xi)}$ of the form \eqref{eq:hybrid_function} such that 
$$
f_{\hat D(\xi)}=\{ f_{\hat A(\xi)}, f_{\hat B(\xi)} \}_H=\{f_{\hat A(\xi)}, f_{\hat B(\xi)}\}_C+ \{ f_{\hat A(\xi)}, f_{\hat B(\xi)}\}_Q.
$$
This property can be proved immediately since the classical bracket $\{f_{\hat A(\xi)}, f_{\hat B}(\xi)\}_C$ defines a function which is quadratic in the quantum degrees of freedom while the quantum bracket  $\{ f_{\hat A(\xi)}, f_{\hat B(\xi)}\}_Q$ is still linear. Just those hybrid magnitudes which depend only on the classical or on the quantum degrees of freedom are closed under the bracket.  After a straightforward computation, and choosing Darboux coordinates $\xi=(\vec q, \vec p)=\{(q^k, p_j)\}_{k,j=1, \cdots, n}$ for $M_C$ we obtain that:
\begin{equation}
\label{eq:classical2}
\{f_{\hat A(\xi)}, f_{\hat B(\xi)}\}_C= \sum_k \left (f_{\partial_{q^k} \hat A(\vec q, \vec p)}f_{\partial_{p_k} \hat B(\vec q, \vec p)}-f_{\partial_{q^k} \hat B(\vec q, \vec p)}f_{\partial_{p_k} \hat A(\vec q, \vec p)} \right )
\end{equation}
and
\begin{equation} 
\label{eq:quantum}
\{f_{\hat A(\xi)}, f_{\hat B (\xi)}\}_Q= f_{[{\hat A },\hat B](\xi)},
\end{equation}
where 
\begin{equation}
\label{eq:commutator}
[\hat A,\hat B](\xi)=-i(\hat A(\xi)\hat B(\xi)-\hat B(\xi)\hat A(\xi)).
\end{equation}

Therefore, we can write that 
\begin{equation}
\label{eq:hybrid}
\{f_{\hat A(\xi)}, f_{\hat B(\xi)}\}_H= \sum_k \left (f_{\partial_{q^k} \hat A(\vec q, \vec p)}f_{\partial_{p_k} \hat B(\vec q, \vec p)}-f_{\partial_{q^k} \hat B(\vec q, \vec p)}f_{\partial_{p_k} \hat A(\vec q, \vec p)} \right )+ f_{[\hat A,\hat B](\xi)},
\end{equation}
i.e, we obtain a combination of quantum-linear functions and quantum-quadratic ones. Clearly, successive brackets will increase further the quantum-degree.

 From the physical point of view, this relation implies that, even if it is Hamiltonian,  the dynamics on the space of observables can not be restricted to quantum-linear functions of the form of Equation \eqref{eq:hybrid_function}. Nonetheless, hybrid energy is conserved, since $\{ f_{\hat H(\xi)}, f_{\hat H(\xi)}\}_H=0$.

If we average this expression with the density $F_{QC}$, we can write that
\begin{multline}
	\label{eq:average-time-evolu}
\frac {d}{dt}	\langle \hat A(\xi)\rangle (t)=\int_{M_C\times M_Q} d\mu_{QC} (\xi, \rho_\psi)F_{QC} (\xi, \rho_\psi) \{f_{\hat A (\xi)}, f_{\hat H(\xi)}\}_H=  \\
\langle f_{\partial_{q^k} \hat A}f_{\partial_{p_k} \hat H} \rangle -\langle f_{\partial_{q^k} \hat H}f_{\partial_{p_k} \hat A} \rangle- \langle f_{[\hat A,\hat H](\xi)} \rangle.
	\end{multline}

Notice that in this expression the requirement of the extension of the algebra is explicit, since we need the average values of pointwise products of the elements of the algebra. Only the last term belong to it. If we write it as an equation on the density $F_{QC}$, the last term affects only its first moment, i.e.,
\begin{multline}
\label{eq:firstmoment}
\langle f_{[A,H](\xi)} \rangle= \int_{M_C}d\mu_C (\xi)\mathrm{Tr} \left ( \hat \rho(\xi) [\hat A(\xi), \hat H(\xi)]   \right )= \\ \int_{M_C}d\mu_C (\xi)\mathrm{Tr} \left ( [\hat H(\xi), \hat \rho(\xi)]  \hat A(\xi) \right ).
\end{multline} 

But the other two can not be written as a function of the first moment only, and higher order moments are required. Indeed, we can write the average value of the product of the functions $f_{\hat A(\xi)}$ and $f_{\hat B(\xi)}$ as 
\begin{equation}
\label{eq:double}
\langle f_{\hat A (\xi)} f_{\hat B(\xi)}\rangle =\int_{M_C} d\mu_C (\xi) \mathrm{Tr} \left ( \hat \rho^{\otimes 2}(\xi) (\hat A(\xi)\otimes \hat B(\xi)) \right ),
\end{equation}
}
where $\hat \rho^{\otimes 2}(\xi)$ is the second quantum moment of the density $F_{QC}$, defined as 
\begin{equation}
\label{eq:rho2}
\hat \rho^{\otimes 2}(\xi) =\int_{M_Q} d\mu_Q (\rho_\psi) F_{QC} (\xi, \rho_\psi) \rho_\psi \otimes \rho_\psi.
\end{equation}
Analogously, the average value of the product of $k$ functions  {\color{black} $f_{\hat A_j(\xi)}$ is obtained as
\begin{equation}
\label{eq:k-functions}
\langle f_{\hat A_1(\xi)} \cdots  f_{\hat A_n(\xi)} \rangle =\int_{M_C} d\mu_C(\xi) \mathrm{Tr} \left ( \hat \rho^{\otimes k} (\xi)  (\hat A_1(\xi)\otimes  \cdots \otimes \hat A_k (\xi) )   \right ),
\end{equation}}
where 
\begin{equation}
\label{eq:k-moment}
\hat \rho^{\otimes k}(\xi)=\int_{M_Q}d\mu_Q(\rho_\psi) F_{QC}(\xi, \rho_\psi) \overbrace{\rho_\psi\otimes \cdots \otimes \rho_\psi}^k
\end{equation}
represents the $k$--th quantum moments of the distribution $F_{QC}$. 

In conclusion, in order to consider Equation \eqref{eq:hybrid_heisenberg} as a dynamical system on the hybrid algebra, the algebra itself must be enlarged. As each classical bracket increases the order of the quantum degrees of freedom by one, clearly we must consider all polynomial functions in $\rho_\psi$ and arbitrary classical dependence as the minimal Poisson algebra generated by hybrid functions of the form \eqref{eq:hybrid_function}.  At each level, higher and higher quantum moments of the distribution are required to compute the average values of the extended functions.  Indeed, if we consider the series of derivatives of order $k$ with respect to time of the average value $\langle \hat A(\xi)\rangle (t)$, we can write it as a function of different terms depending on all the moments $\hat \rho^{\otimes j} (\xi)$ up to order $k+1$ evaluated on different combinations of $\hat A(\xi)$ and $\hat H(\xi)$:
\begin{equation}
\label{eq:time-k}
\frac{d^k}{dt^k}\langle \hat A(\xi)\rangle (t) =F \left (\hat \rho^{\otimes 0} (\xi), \cdots, \hat \rho^{\otimes k+1} (\xi), \hat A(\xi), \hat H(\xi) \right ).
\end{equation}

{\color{black} Hence, 
for a finite range of time $(t_0, t_0+\Delta t)$, the behavior of an average value $\langle \hat A(\xi) \rangle (t)$ can be approximated to arbitrary precision by a finite number of quantum moments of the distribution $F_{QC}$.}
Notice that the particular time scale and the precision depends on the particular observable $\hat A(\xi)$. In any case, by considering a finite time range we see that the concept of hybrid observable changes. Time introduces correlations between the classical and quantum subsystems and the nonlinear evolution makes that correlation measurable in the time-dependence of the average values. We can see that in Equation \eqref{eq:average-time-evolu}. If we want to consider a linearized time dependence in $\langle \hat A(\xi) \rangle$, we need to compute average values of pairs of quantum operators\footnote[1]{Notice that the analysis done here for time evolution can be repeated for any transformation requiring to introduce the necessary infinitesimal generator, that in this case is the Hamiltonian. This justifies extending the hybrid algebra to products of arbitrary operators, as any operator can be used to define a transformation of the system.}. The time range which can be approximated by a finite number of time derivatives may depend on the observable we consider.  But for any observable, there will be a certain time scale where the behavior of the system can be approximated to any precision with a sufficiently high order of quantum moments. As the application of our model will be a numerical simulation of the system at a certain time scale, this finite number of moments should be enough to characterize the state of the system. In the following sections we will learn to write the dynamics of the quantum moments and thus the dynamics of the physical macrostate for the relevant time range.

\section{Hybrid states and dynamics}
\label{sec:Hybrid_states}

\subsection{Writing a dynamics for the physical states}
Let us consider now how to define an equation to write the {\color{black} evolution of the average value $\langle \hat A(\xi) \rangle$ given by \eqref{eq:average-t} } as an evolution equation on the physical state. 
  Following \cite{balescuStatisticalDyamicsMatter1997} we can associate a {\color{black} Liouville} equation to the probability density and define the curve on the space of probability densities which reproduces the evolution of the average values $\langle \hat A(\xi)\rangle (t)$, i.e.,{\color{black}
\begin{equation}
\label{eq:averages2}
\langle \hat A(\xi)\rangle (t)=\int_{M_C\times M_Q} d\mu_{QC} (\xi, \rho_\psi) F_{QC} (\xi, \rho_\psi; t) f_{\hat A} (\xi),
\end{equation}
where $F_{QC} (\xi, \rho_\psi; t)$ is the solution of the master equation. In \cite{alonsoStatisticsNoseFormalism2011} we proved that because the dynamics is Hamiltonian we can obtain a dynamical equation for the density $F_{QC}$ as the Liouville equation 
\begin{equation}
\label{eq:Liouville}
\frac{d F_{QC}}{dt}=\{ f_{\hat H(\xi)}, F_{QC} \}_H.
\end{equation}}
This equation translates the microstate dynamics to the full probability density. Nonetheless, if we are interested in the behavior of average values of physical magnitudes for a certain time scale, we know that not all the density is required, only a few of its quantum moments. Hence, we are going to study now how can we write the effect of Ehrenfest dynamics on those moments.

\subsection{The initial problem: the dynamics of the first quantum moment}
Let us consider now  {\color{black} the first moment of the distribution $F_{QC}$, i.e., the family of operators $\hat \rho(\xi)$.  We can consider Liouville equation for the density and re-write it in terms of the first moment:
\begin{multline}
  \label{eq:rhodot1}
  {\color{black}\frac{d{\hat{\rho}} (\xi)}{dt}} = \int_{M_{Q}} \mathrm{d}\mu_{Q} (\rho_\psi)\{f_{H}, F_{QC}\}_H(\xi, \rho_{\psi}) \rho_{\psi}= \\ [\hat H(\xi),  \hat \rho(\xi)] + \int_{M_{Q}} \mathrm{d}\mu_{Q}(\rho_\psi) \{f_{H}, F_{QC}\}_C(\xi, \rho_{\psi}) \rho_{\psi}.
\end{multline}
As the time dependence of average values can be written as a result of dynamics written on the space of operators or on the space of states,
\begin{equation}
  \langle \hat A \rangle (t)=
  \begin{cases}
    \int_{M_{C}} \mathrm{d} \mu_{C} (\xi)\mathrm{Tr}[\hat{\rho}(\xi;t) \hat{A}(\xi)] \\
    \int_{M_{C}} \mathrm{d} \mu_{C}(\xi) \mathrm{Tr}[\hat{\rho}(\xi) \hat{A}(\xi; t)]
  \end{cases},
\end{equation}
Equation \eqref{eq:rhodot1} must satisfy
\begin{equation}
  \frac{\mathrm{d}}{\mathrm{d}t} \langle \hat{A}(\xi) \rangle = \int_{M_{C}} \mathrm{d} \mu_{C} (\xi)\mathrm{Tr}[\dot{\hat{\rho}}(\xi) \hat{A}(\xi)]=-\int_{M_{C}} \mathrm{d} \mu_{C}(\xi) \mathrm{Tr}[\hat{\rho}(\xi) \dot{\hat{A}}(\xi)],
\end{equation}
where the dot represents the tangent vector of each respective curve (on states or on operators). Notice that, by definition, 
\begin{equation}
  \label{eq:rho_t}
  \hat{\rho}(\xi;t):=\int_{M_Q}d\mu_q(\rho_\psi) F_{QC}(\xi, \rho_\psi; t) \rho_{\psi}.
\end{equation}
Therefore, we can conclude that the properties of the first quantum moment will be preserved (normalization, positivity, etc), if the solution of the Liouville equation defines a curve of well-defined probability densities.

As we saw in the previous section, the set of linear functions of the form  $f_A$ is not closed under the hybrid Poisson bracket. This has an important impact on the Liouville equation \eqref{eq:Liouville}. Indeed, a master equation for the density $F_{QC}$ which captures the dual behavior of the dynamics of the hybrid observables (which does not have linear functions as a Poisson subalgebra)} can not be restricted to the first quantum moment, which is only able to capture the linear (quantum) functions of the form \eqref{eq:hybrid_function}. If we must include quantum polynomial functions of orders higher than one {(\color{black} i.e., objects of the form $f_Af_H$)}, higher quantum moments of the distribution $F_{QC}$ must be considered. But then the equivalence relation given by the first moment is broken: two measures having identical first quantum moments may evolve in different ways and produce different hybrid states for each value of time{\color{black}, depending on the higher quantum moments}.  Let us consider this issue in some detail in a particular example.

{\color{black}
\begin{example}
As we saw above, there exist many densities $F_{QC}$ with the same first quantum moment. Let us see now that the nonlinearity of Ehrenfest equations makes impossible to write Ehrenfest dynamics in terms $\hat \rho(\xi)$ in a consistent way.  

Let us now construct a particular hybrid system with Ehrenfest dynamics, with an ad-hoc Hamiltonian and density matrix. We will then find a one-parameter family of different densities with the same first moment $\hat{\rho}(\xi)$ and see that the tangent vector defined by each density is different.  More in particular:
     \begin{enumerate}
       \item We consider one classical degree of freedom, $M_{C} \equiv \mathbb{R}^{2}$ and $\xi \equiv \left(R, P \right)$. 
       \item We choose a two-level system as the quantum subsystem. Therefore, the space of projectors is:
       \[
       M_{Q} = \{ \rho_{\psi} \mid \rho_{\psi} \in \mathrm{Herm}(2) = \mathrm{span}_\mathbb{R} \{ \sigma_{0}, \sigma_{1}, \sigma_{2}, \sigma_{3} \}, \rho_{\psi}^{2} = \rho_{\psi}, \mathrm{Tr}(\rho_{\psi}) = 1 \}
       \]
       with $\{\sigma_{i}\}_{i = 1}^{3}$ being the Pauli matrices and $\sigma_{0}$ the identity. The usual description of this space in quantum mechanics is \textit{the Bloch sphere}. It can be proved that $M_{Q}$ as defined above is bijective to $S^{2} = \{x^{2} + y^{2} + z^{2} = 1 \mid x,y,z \in \mathbb{R}\} \subset \mathbb{R}^{3}$, with coordinates $\{x = \mathrm{Tr}[\sigma_{1}\rho_{\psi}], \; y = \mathrm{Tr}[\sigma_{2}\rho_{\psi}], \; z = \mathrm{Tr}[\sigma_{3}\rho_{\psi}]\}$, which will be called \textit{Bloch coordinates}, denoted by a subscript $\mathcal{B}$.
       \item The Hamiltonian of the system is defined as
       \begin{equation}
        \label{eq:Hamiltonian_example}
        \hat{H}(\xi) = \hat{H}(R,P) = \frac{1}{2}(R^{2} + P^{2}) \mathbb{I}_{2} + E_{1}(R,P) \hat{\pi}_{1}(R,P) + E_{2}(R,P)\hat{\pi}_{2}(R,P)  
       \end{equation}
       where $\hat{\pi}_{k} \in M_{Q}$ are classical-point-dependent projectors: $\hat{\pi}_{1}(R,P) = \left( \: \sin R, 0, \cos R \: \right)_{\mathcal{B}}$ and $\hat{\pi}_{2}(R,P) = \left( \: -\sin R , 0, -\cos R \: \right)_{\mathcal{B}}$. Note that they are orthogonal by construction. The two energy levels $E_{k}$ are:
       \[
       E_{1}(R,P) = \frac{1}{1 + R^2} \qquad E_{2}(R,P) = E_{1}(R,P) + 1 + 0.1 R^2 \:.
       \]
       
       \item The hybrid density matrix is defined as
       \[
       \hat{\rho}(\xi) = F_{C}(\xi)\lambda(\xi)\:\hat{\pi}_{1}(\xi) + F_{C}(\xi)\left(1 - \lambda(\xi)\right)\:\hat{\pi}_{2}(\xi)
       \]
with $\lambda(R,P) = \frac{2}{\pi}\operatorname{atan}{\left(R^{2} + P^{2} \right)}$, $F_C(R,P) = \frac{1}{2 \pi} \exp\left({- \frac{R^{2} + P^2}{2}}\right)$ and with projectors defined as, $\hat{\pi}_{1}(R,P) = \left( \: \sin a, 0, \cos a \: \right)_{\mathcal{B}}$ and $\hat{\pi}_{2}(R,P) = \left( \: -\sin a, 0, -\cos a \: \right)_{\mathcal{B}}$, where $a = \operatorname{atan}{\left(R^{2} + P^{2} \right)}$.

     \end{enumerate}

Finally, we define a family of densities $F_{QC}^\theta$ in terms of the arbitrary parameter $\theta \in \mathbb{R}$.  Consider a decomposition of the matrix $\hat \rho(\xi)$ as sum of projectors 
       \begin{equation}
        \hat \rho(\xi)=\sum_k \lambda_k(\xi) \hat \pi_k(\xi), \qquad \hat \pi_k(\xi)^2=\hat \pi_k(\xi).
       \end{equation}
  It is immediate to prove that the density
  \begin{equation}
    F_{QC}(\xi, \rho_\psi)=\sum_k \lambda_k(\xi) \delta(\rho_\psi-\hat \pi_k(\xi)),
  \end{equation}
  has $\hat \rho(\xi)$ as first moment. We can consider a family of the decompositions choosing the coordinates of one projector to be $(\sin \theta, 0, \cos \theta)_{\mathcal{B}}$, and solving the geometric problem in the Bloch sphere to find the other projector and the corresponding weights. Thus, we construct the $\theta$-dependent family of densities $F_{QC}^\theta$ having the same first quantum moment.  For each density, consider Equation \eqref{eq:rhodot1} and represent the corresponding tangent vector as a function of the parameter $\theta$. 
     We present graphically the results in Fig. \eqref{fig:fig1}. We observe that the $x$ and $z$ coordinates of $\dot{\hat{\rho}}(\xi)$ (in the Bloch basis) depend of the arbitrary parameter.
     
     \begin{figure}[h] 
       \centering
       \begin{subfigure}{0.49\textwidth}
        \includegraphics[width=\textwidth]{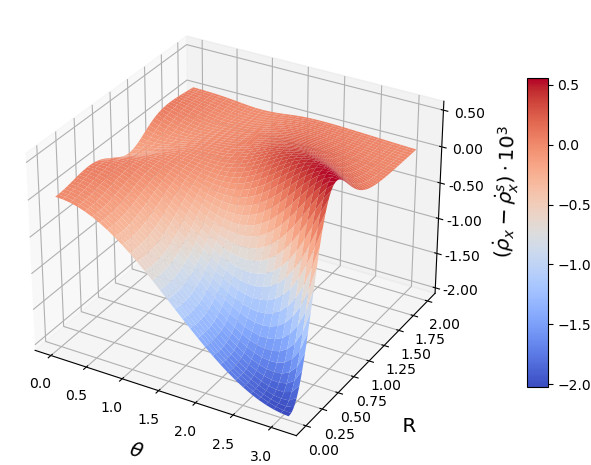}
        \caption{$x$-coordinate, $\frac{1}{2}\mathrm{Tr}(\dot{\rho}(\xi)\sigma_{1})$.}
       
       \end{subfigure}
       \begin{subfigure}{0.49\textwidth}
        \includegraphics[width=\textwidth]{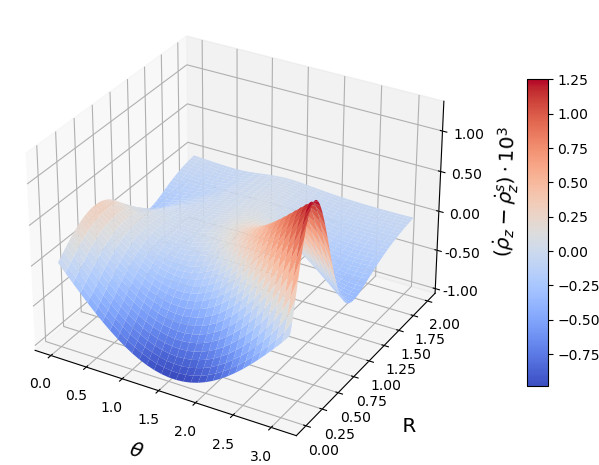}
        \caption{$z$-coordinate, $\frac{1}{2}\mathrm{Tr}(\dot{\rho}(\xi)\sigma_{3})$.}
       
       \end{subfigure}
       \caption{Difference of $x$ and $z$ components of $\dot{\rho}(\xi)$ for different values of the classical parameter $R$ and $\theta$.}
       \label{fig:fig1}
     \end{figure}
As expected, we verify that the dynamics of the first quantum moment depend on the parameter $\theta$ and therefore that the dynamics of the first moment is not well defined.
Hence, we see that different density dynamics are possible for a given $\hat \rho(\xi)$, depending on elements which are not physically observable, \textit{a prioiri}, at the (quantum) linear level.
Notice that the origin of the problem is the nonlinearity of Ehrenfest dynamics, which defines a different evolution for the two different quantum states coupled to the same classical point $\xi$. Had we considered a pure quantum system with unitary evolution, or even an uncoupled hybrid system,  the evolution would be well defined because of the linearity of the pure quantum evolution.
Let us see now how the dynamics can be determined when considering higher order moments.
    \end{example}
}

\subsection{The dynamics of higher quantum moments}
\label{sec:nonlinear_master_equations}

We have seen above how at the linear level, we have part of the information of the full Liouville equation, although not all. Nonetheless, Equation \eqref{eq:rhodot1} is enough to read the behavior of the two marginal distributions $F_C(\xi)=\mathrm{Tr} \hat \rho(\xi)=\int_{M_Q}d\mu_Q F_{QC}(\xi, \rho_\psi)$ and $\hat \rho=\int_{M_C} d\mu_C(\xi) \hat \rho(\xi)$.  Indeed, marginalizing one of the variables makes one of the two brackets to vanish, and that produces:
\begin{equation}
  \frac{d}{dt} F_C(\xi)=\mathrm{Tr}\left(\{\hat H(\xi),\hat\rho(\xi)\}_C\right),
  \end{equation}
  and
  \begin{equation}
    \frac{d}{dt} \hat\rho=\int_{M_C}d\mu_C(\xi)[\hat H(\xi),\hat\rho(\xi)].
  \end{equation}

These two equations coincide with those obtained from different approaches to hybrid quantum-classical dynamics, such as \cite{aleksandrovStatisticalDynamicsSystem1981,Kapral1999}, where Ehrenfest equations were not considered. This is a remarkable result, since it ensures that, at least at first order, the evolution of pure classical or pure quantum magnitudes behave the same as in those other approaches even for coupled quantum-classical systems.  

What about the full hybrid (i.e., non-marginalized) behavior? Can we read any useful properties from Liouville equation? Indeed, we can. 
We saw above that writing Ehrenfest dynamics on the physical states requires of higher order quantum moments to be able to capture the effects of the nonlinearity and the non-closeness of the Poisson algebra. Dynamics written on the first moment only is not well defined, as the nonlinearity of the dynamics makes it dependent on the choice of the decomposition of the quantum operator at each classical point.  In this section, we will see that the dynamics of each moment depends on the next one; the Liouville equation being thus equivalent to an infinite series of differential equations, one for each $\hat \rho^{\otimes k}(\xi)$. 

Let us consider again the definition of the $k$--th moment given in Equation \eqref{eq:k-moment}. By construction, they are operators on the Hilbert space $\bigotimes_{j=1}^k\mathcal{H}_j$. We can define partial traces if we average over some of those copies, since $\mathrm{Tr} \rho_\psi=1$ and hence $\mathrm{Tr}_p \overbrace{\rho_\psi\otimes \cdots \otimes \rho_\psi}^k=\overbrace{\rho_\psi\otimes \cdots \otimes \rho_\psi}^{k-p}$ for $p\leq k$. It is immediate then that 
\begin{equation}
\label{eq:partial_traces}
\mathrm{Tr}_p \hat \rho^{\otimes k}(\xi)=\hat \rho^{\otimes k-p}(\xi),
\end{equation}
where the zeroth order moment corresponds to the marginal classical density
\begin{equation}
\label{eq:order0}
\mathrm{Tr}_k \hat \rho^{\otimes k}(\xi)=\hat \rho^{\otimes 0}(\xi)=\int_{M_Q}{\color{black}d\mu_Q} (\rho_\psi) F_{QC}(\xi, \rho_\psi)=F_C(\xi).
\end{equation}
{\color{black} Despite of the symmetric nature of the expressions involving the points $\rho_\psi$ in the following we will consider traces of different operators, and will denote as $\mathrm{Tr}_1$ the trace acting always on the first one
\begin{equation}
\label{eq:trace-sym}
\mathrm{Tr}_1 (A\otimes B)= \mathrm{Tr}(A)B .
\end{equation}
}
In an analogous way, the expression is extended to arbitrary traces $\mathrm{Tr}_k$.

Using these relations, it is immediate to re-write Equation \eqref{eq:average-time-evolu} as 
\begin{multline}
\label{eq:av-evolu-rho}
\frac {d}{dt}	\langle \hat A(\xi)\rangle (t)=
\langle f_{\partial_{q^k} A}f_{\partial_{p_k} H} \rangle -\langle f_{\partial_{q^k} H}f_{\partial_{p_k} A} \rangle- \langle f_{[A,H]} \rangle = \nonumber \\
\int_{M_C}d\mu_C(\xi)\mathrm{Tr}([\hat A,\hat H]\hat\rho(\xi))
    +\int_{M_C}d\mu_C(\xi) \sum_k\mathrm{Tr}\left ( \left (\partial_{[q^k}\hat A \otimes \partial_{p_k]}\hat H \right )\hat \rho^{\otimes 2}(\xi) \right ),
\end{multline}
where $\partial_{[q^k}\hat A \otimes \partial_{p_k]}\hat H= \partial_{q^k}\hat A(\xi) \otimes \partial_{p_k}\hat H(\xi)-\partial_{q^k}\hat H(\xi) \otimes  \partial_{p_k}\hat A(\xi)$. 

From that expression, and using the compact support of the classical functions, we can read the time derivative of the first quantum moment as 
\begin{equation}
\label{eq:diff_rho}
\dot {\hat \rho} (\xi)= [\hat H(\xi), \hat \rho(\xi)]+ \mathrm{Tr}_1 \left (  \left \{ {\color{black}\hat H(\xi) \otimes \mathbb{I}} , \hat \rho^{\otimes 2} (\xi)\right \}_C  \right ),
\end{equation}
where $\dot {\hat \rho} (\xi)$ is defined by Equation \eqref{eq:rhodot1}. Analogously, we can define the time derivative of any moment $\dot {\hat \rho}^{\otimes k} (\xi)$ as a function of ${\hat \rho}^{\otimes k}(\xi)$ and ${\hat \rho}^{\otimes k+1}(\xi)$.

\begin{equation}
\label{eq:diff_rhok}
\dot {\hat \rho}^{\otimes k} (\xi)= [\hat H^k(\xi), \hat \rho ^{\otimes k}(\xi)]+ \mathrm{Tr}_1 \left (  \left \{ {\color{black}\hat H(\xi) \otimes \overbrace{ \mathbb{I}\otimes \cdots \otimes \mathbb{I}}^{k} }, \hat \rho^{\otimes k+1} (\xi)\right \}_C  \right ),
\end{equation}
where 
{\color{black}
\begin{equation}
\label{eq:Hk}
\hat H^k= \overbrace{\hat H(\xi)\otimes \mathbb{I}\otimes \cdots \otimes \mathbb{I}}^{k} +
\cdots +\overbrace{\mathbb{I}\otimes \cdots \otimes \mathbb{I}\otimes\hat H(\xi) }^{k}
\end{equation}
}

In this way, we find a system of differential equations for all the quantum moments $\{ {\hat \rho}^{\otimes k}(\xi)\}_{k=0, \cdots, }$ which is equivalent to the Liouville equation for the density $F_{QC}$ (Equation \eqref{eq:Liouville}).

\subsection{An effective equation}

{\color{black}As we saw above, only a finite number of quantum moments are necessary to recover the behavior of the average value of physical observables for finite time intervals. Notice, though, that when we make the approximation for a certain time range, the approximated solution of the dynamics of the hybrid observables $\hat A(\xi; t)$ defines a different dynamics for the average  values $\langle \hat A(\xi) \rangle (t)$. As in each case the average values involve a finite number of quantum moments, we can define  different dynamics on the space of states with the corresponding Equation \eqref{eq:diff_rhok}.} But there is a problem to do that: the dynamics of the $k$--th quantum moments, depends on the $k+1$--th one and therefore it is not possible to solve the equation for a certain $k$ without the solution to the next level.  Nonetheless, as the time scale is only able to capture the $k$-first moments, it is not possible to do any measurement or sequence of measurements {\color{black} which depend on those degrees of freedom.  Hence, as the $(k+1)$--th moment can not be known, we may consider an effective dynamical equation for $\hat \rho^{\otimes k}(\xi)$ where  the value of $\hat \rho^{\otimes k+1}(\xi)$} represents the minimum knowledge on the system compatible with the physical constraints. In order to do that, we must introduce a suitable notion of entropy.

In \cite{alonsoEntropyCanonicalEnsemble2020} we introduce a notion of hybrid entropy for the first quantum moment $\hat \rho(\xi)$. We built it based on the bivariate distribution that the first moment represents, and the corresponding factorization in marginal and conditional probabilities.  That factorization makes sense for any of the quantum moments $\hat \rho^{\otimes k}(\xi)$, which, on the other hand, happens to be formally analogous to $\hat \rho(\xi)$ with the only difference of being an operator on the product Hilbert space $\bigotimes_{j=1}^k \mathcal{H}$:
\begin{itemize}
  \item the marginal probability of the bivariate probability distribution represented by $\hat \rho^{\otimes k}(\xi)$ corresponds to the classical density 
  \begin{equation}
  \label{eq:classical_marginal}
  F_C(\xi)=\mathrm{Tr}_k \hat \rho^{\otimes k}(\xi)=\int_{M_Q}d\mu_Q(\rho_\psi) F_{QC}(\xi, \rho_\psi).
  \end{equation}
  \item  The corresponding quantum conditional probability $\mathcal{P}(\rho_\psi \mid \xi)$ is the probability of a pure quantum system (the classical degrees of freedom  $\xi$ are fixed). As such, Gleason theorem (\cite{gleasonMeasuresClosedSubspaces1957}) ensures that a well-defined density matrix  $\hat \rho_\xi^{\otimes k}$ (i.e., a trace class, normalized and non-negative self-adjoint operator on $\bigotimes_{j=1}^k \mathcal{H}$) must exist to represent it.  
\end{itemize}
This representation allows to implement the exclusivity of hybrid events in a simpler way, since it incorporates automatically the orthogonality of the mutually exclusive quantum events (as the usual density matrix does). See \cite{alonsoEntropyCanonicalEnsemble2020} for a more detailed discussion.

By using this factorization, we can adapt the construction of the hybrid entropy in \cite{alonsoEntropyCanonicalEnsemble2020}. As the entropy for the state $\hat \rho^{\otimes k}(\xi)$ must be written as the sum of the entropy of the marginal distribution \eqref{eq:classical_marginal} and the classical average of the entropy of the conditional one we obtain
\begin{multline}
\label{eq:entropy}
S\left [ \hat{\rho}^{\otimes k}(\xi)\right ]=  \\ -k_B\int_{M_C} d\mu_C (\xi) F_C(\xi) \log F_C(\xi)  - k_B
\int_{M_C}d\mu_C(\xi) F_C(\xi) \mathrm{Tr}_k \left (  \hat{\rho}_\xi^{\otimes k} \log \hat {\rho}_\xi^{\otimes k} \right )  \\ 
= -k_B\int_{M_C}d\mu_C(\xi) \mathrm{Tr}_k \left (  \hat{\rho}^{\otimes k}(\xi) \log \hat {\rho}^{\otimes k}(\xi) \right ).
\end{multline}

This function represents the entropy of the hybrid system modelled by the $k$--th quantum moment. Using this entropy function, we may consider searching for the state $\hat \rho^{\otimes k+1}(\xi)$ which maximizes the entropy subject to some constraints to use in the right hand side of  Equation \eqref{eq:diff_rhok}. The natural constraint is asking this first unknown moment to produce the last known one by trace, i.e., we will search for a $k+1$ moment such that
\begin{equation}
\label{eq:tr1}
\mathrm{Tr}_1 \hat \rho^{\otimes k+1}(\xi)= \hat \rho^{\otimes k}(\xi),
\end{equation}
where $\hat \rho^{\otimes k}(\xi)$ is the moment whose dynamics is being defined. The unknown components of the moment are thus only the degrees of freedom being integrated out by the trace. As it is not possible to determine that information by any measurement, we select the state with the maximum possible entropy for those degrees of freedom to act as a source for the dynamical equation. In the following we shall represent as $\hat \rho^{\otimes k}_{\mathrm{MaxEnt}} (\xi)$  the state which maximizes the entropy with constraint \eqref{eq:tr1}.

Notice that we are using the MaxEnt formalism from the point of view of information {\color{black}(see \cite{jaynesInformationTheoryStatistical1957})}, and not as it is usually done when trying to identify thermodynamical ensembles. Those thermodynamical problems can also be considered, though.  The entropy function would be the same, but depending on the problem, other constraints would be considered. For instance, we may study the state which maximizes entropy \eqref{eq:entropy}  while keeping the normalization and the average value of the Hamiltonian $\hat H(\xi)$ fixed. This would produce the candidate to model the canonical ensemble in our context, as we considered in \cite{alonsoEntropyCanonicalEnsemble2020} for the case of the first moment.
In a similar way we may consider a microcanonical ensemble $\hat \rho^{\otimes k}_{MCE} (\xi)$. All these situations will be considered in detail in a forthcoming paper.

We conclude thus that this MaxEnt formalism with constraint \eqref{eq:tr1} provides us with a natural candidate to represent the state $\hat \rho^{\otimes k+1}(\xi)$ in Equation \eqref{eq:diff_rhok}. As there is no physical measurement which can inform us about the state, we consider it to be the state which  maximizes the uncertainty about it, which is precisely $\hat \rho^{\otimes k+1}_{\mathrm{MaxEnt}} (\xi)$. Therefore, we can write as effective dynamics for the $k$--th quantum moments which is the equation:
\begin{equation}
\label{eq:rho_k_effective}
\dot {\hat \rho}^{\otimes k} (\xi)= [\hat H^k(\xi), \hat \rho ^{\otimes k}(\xi)]+ \mathrm{Tr}_1 \left (  \left \{ {\color{black}\hat H(\xi) \otimes \overbrace{ \mathbb{I}\otimes \cdots \otimes \mathbb{I}}^{k}}
 , \hat \rho^{\otimes k+1}_{\mathrm{MaxEnt}} (\xi)\right \}_C  \right ). 
\end{equation}
This equation is well defined, since it only depends on the degrees of freedom of the $k$--th moment (given that the constraint \eqref{eq:tr1} must hold for all times) and can be used to approximate the dynamics of the average values of physical magnitudes in a finite range of times for statistical systems whose microstates follow Ehrenfest dynamics. 

{\color{black}
\begin{example}
In order to provide a practical application of our framework, 
let us consider again the simple example considered above of a system with two dimensional classical phase space and a two level quantum system. Let us assume, for simplicity, that we just consider Equation \eqref{eq:rho_k_effective} the first quantum moment. In that case, we have to identify the solution of Maximal Entropy for the case $k+1=2$. Let us proceed.

First of all, we are going to consider the factorization in marginal and conditional probabilities given by Equation \eqref{eq:factorization_rhoxi}, for $\hat \rho(\xi)$ and ${\hat \rho}^{\otimes 2}_\xi$:
\begin{equation}
  \label{eq:factorization}  
  \hat \rho(\xi)=F_C(\xi)\hat \rho_\xi; \qquad \hat \rho^{\otimes 2}(\xi)=F_C(\xi)\hat {\rho^{\otimes 2}}_\xi.
\end{equation}
By construction, $\hat\rho_\xi$ is written as a combination of the basis $\{ \hat \sigma_j\}_{j=0, 1,2, 3}$, and therefore  $\hat {\rho^{\otimes 2}}_\xi$  must be written as a combination of the 10-dimensional symmetrical basis $\{ \frac 12 (\hat \sigma_j \otimes \hat \sigma_k + \hat \sigma_k \otimes \hat \sigma_j) \}_{j\leq k=0,1,2,3}$. The corresponding expressions read:
\begin{align}
\label{eq:rho1-2}
\hat \rho_\xi =&\sum_{j=0}^3  \mu_j \hat \sigma_j; \\ 
{\hat \rho}^{\otimes 2}_\xi =& \mu_{00} \hat \sigma_0\otimes \hat \sigma_0+ \frac 12 \sum_{k=1}^3 \mu_{0k} (\hat \sigma_0\otimes \hat \sigma_k + \hat \sigma_k \otimes \hat \sigma_0)\\ &+\frac 12 \sum_{j\leq k=1}^3 \mu_{jk} (\hat \sigma_j\otimes \hat \sigma_k + \hat \sigma_k \otimes \hat \sigma_j),
\end{align}
where we can relate these coordinates with the average values with respect to the conditional distribution $F_\xi^{cond}$ of the coordinate functions of the points of $M_Q$ as:
\begin{equation}
  \label{eq:average}
\mu_j=\mathbb{E}(\mu_j(\rho_\psi)); \qquad  \mu_{jk}=\mathbb{E}(\mu_j(\rho_\psi)\mu_k(\rho_\psi)),
\end{equation}
$\mu_j(\rho_\psi)$ representing the $j$--th coordinate of the pure state $\rho_\psi$ with respect to the basis $\{\hat \sigma_j \}$.

Furthermore, as $\hat \rho_\xi$ and $\hat \rho_\xi^{\otimes 2}$ are density matrices, their trace must be normalized. This implies that
\begin{equation}
  \label{eq:normalization}
  \mu_0=\frac 12, \qquad \mu_{00}=\frac 14.
\end{equation}
 
In this context, constraint \eqref{eq:tr1} implies that
$$
2 \mu_{0j}=\mu_j, \quad j= 1,2, 3.
$$
Notice, though, that from the definition of $\hat \rho^{\otimes 2}(\xi)$ there are also a few relations between the coordinates,  to be fulfilled. In particular
\begin{itemize}
 \item As $\hat \rho_\xi$ must be a well defined density matrix, we know that its purity must be lower or equal to one, i.e.
 \begin{equation}
  \label{eq:condi2}
  \sum_j \mu_j^2\leq \frac 12.
 \end{equation}
 \item An analogous property holds for ${{\hat \rho}^{\otimes 2}}_\xi$: its purity must also be lower than one, i.e.,:
 \begin{multline}
  \label{eq:condi3}
  4 \left ( \mu_{00}^2 + \frac 12 \sum_k \mu_{0k}^2+ \frac 12 \sum_{i\leq j} \mu_{ij}^2     \right ) \leq 1 \Rightarrow 
  \sum_k \mu_{0k}^2+ \sum_{i\leq j} \mu_{ij}^2 \leq \frac 38.
 \end{multline}

   \item  As $\mu_{jj}$ is the expectation value of the square of the function $\mu_j(\rho_\psi)$ on $M_Q$  we can easily verify that  
   \begin{equation}
    \label{eq:condi4}
    \mu_{jj}-\mu_j^2=\mathbb{E}(\mu_j^2(\rho_\psi))-\mathbb{E}(\mu_j(\rho_\psi))^2=\mathbb{E}\left ((\mu_j(\rho_\psi) -\mu_j)^2\right ) \geq 0,
   \end{equation} 
   since the variance of the function $\mu_j(\rho_\psi)$ must be positive definite for a well-defined probabilistic system.
   
   \item As points $\rho_\psi\in M_Q$ are pure states, $\sum_j \mu_j^2(\rho_\psi)=\frac 12$ and therefore
\begin{equation}
  \label{eq:pure-states}
  \sum_{j=1}^3 \mu_{jj}=\mathbb{E}\left ( \sum_j \mu_j^2(\rho_\psi)   \right )=\mathbb{E}\left  (\frac 12 \right )=\frac 12.
\end{equation}
   \item Analogously, we can verify that, from Cauchy-Schwartz inequality 
   \begin{equation}
    \label{eq:condi5}
    \vert\mu_{ij}-\mu_i\mu_j\vert\leq \sqrt{(\mu_{ii}-\mu_i^2)(\mu_{jj}-\mu_j^2)} 
   \end{equation}
   
\end{itemize}

With these constraints, we search for the MaxEnt solution for the hybrid entropy function \eqref{eq:entropy}. The result defines a relation between the free variables of $\hat {\rho}^{\otimes 2}_\xi$ with the other variables, i.e. which depend on those of $\hat {\rho}_\xi$:
\begin{equation}
\mu_{jk}=\mu_{jk} (\mu_{l}); \qquad j,k=1,2,3, \quad l=1,2,3.
\end{equation}

Computing entropy involves a complicated expression of the logarithm of $\hat \rho^{\otimes 2}(\xi)$ (or, analogously, of  $\hat \rho_\xi^{\otimes 2}$ if we use the marginal-conditional factorization). For the sake of simplicity, instead computing the spectrum of the matrix, we can use Mercator series to approximate it by a polynomial of the traces of the powers of the matrix, i.e.
\begin{multline}
  \label{eq:Mercator-entropy}
  S^{cond}=-\mathrm{Tr}(\hat \rho_\xi^{\otimes 2} \log \hat \rho_\xi^{\otimes 2})\simeq \mathrm{Tr} 
  \left ( \hat \rho_\xi^{\otimes 2} \left ( \left (\mathbb{I}- \hat \rho_\xi^{\otimes 2}\right) +\frac 12 \left (\mathbb{I}- \hat \rho_\xi^{\otimes 2}\right)^2 +\cdots \right )   \right )
\end{multline}

Keeping only the linear approximation of the entropy, we would obtain
\begin{equation}
  \label{eq:linear}
  S^{cond}\simeq 1- \mathrm{Tr}\left (\hat \rho_\xi^{\otimes 2} \right )^2=1- 
  4 \left ( \mu_{00}^2 + \frac 12 \sum_k \mu_{0k}^2+ \frac 12 \sum_{i\leq j} \mu_{ij}^2     \right ).
\end{equation}

Maximizing this function in the region compatible with the constraints above, considering the symmetry between the different group of indices, can be achieved by fixing:
\begin{equation}
  \label{eq:solution}
  \begin{cases}
  \mu_{0k}= \frac{\mu_k}{2} \text{ for }  k=1,2,3 \\
  \mu_{jk}= 0 \text{ for } j\neq k=1,2,3 \\
  \mu_{kk}=\mu_k^2+ \frac 13 \left ( \frac 12-\mathcal{P}(\hat \rho_\xi)  \right ) \text{ for }  k=1,2,3 
  \end{cases}
\end{equation}
where $\mathcal{P}(\hat \rho_\xi)=\sum_j \mu_j^2$ represents the purity of the state $\hat \rho_\xi$. This point can be considered just a simple approximation to the real MaxEnt solution, but it helps us to illustrate how our construction works. Higher order approximations can be implemented with numerical tools on the expansion above.

With the solution of the MaxEnt problem we can write  equation \eqref{eq:rho_k_effective} becomes, written in terms of the coordinates as:
\begin{equation}
  \label{eq:ODE-0}
  \dot \mu_0(\xi)= \sum_{k=0}^3 \{ H_k(\xi), \mu_{k}(\xi)  \}_C
\end{equation} 

\begin{equation}
  \label{eq:ODE-j}
  \dot \mu_j(\xi)= \sum_{kl}c_{kl}^j \mu_k(\xi) H_l(\xi) + \frac 12\{H_0(\xi), \mu_j(\xi)\}_C+ \sum_{k=1}^3\{ H_k(\xi), \mu_{jk}(\xi)  \}_C, \quad j=1, 2,3
\end{equation} 
where we used that $c_{kl}^j$ represent the structure constants of $\mathfrak{u}(2)$ in the basis we chose, $\mu_j(\xi)=F_C(\xi)\mu_j$ and $\mu_{jk}(\xi)=F_C(\xi)\mu_{jk}$. 

We can see in these equations how classical and quantum degrees of freedom evolve coupled, the marginal classical density $\mu_0(\xi)=F_C(\xi)$ depending on the quantum degrees of freedom $\mu_j(\xi)$, whose evolution is also governed by $\mu_0(\xi)$.  Furthermore, from the expressions of Equations \eqref{eq:solution} and \eqref{eq:ODE-0} and \eqref{eq:ODE-j}, it is immediate to see that the final dynamics of $\hat \rho(\xi)$ is non-linear.
\end{example}
}

\section{Conclusions}
\label{sec:Conclusions}

In this paper we have considered the problem of dynamical statistical systems for hybrid quantum-classical systems having Ehrenfest dynamics as microstate dynamics.  We have seen how dynamical evolution (or any other type of observable-generated transformation) enlarges the set of physically observable magnitudes since it introduces correlations between the magnitudes of the form $\hat A(\xi)$ which are the natural models of hybrid magnitudes. The nonlinearity introduced by the classical subsystem is responsible of that correlation, which produces a progressive enlargement of the hybrid subalgebra. This implies that, while (linear) hybrid observables depend only on the first quantum moment of the distribution,  higher quantum moments are required to compute the evolution of the average value of physical observables. For a given time interval and a certain accuracy, only a finite number of those quantum moments are required to approximate the evolution of the average values of arbitrary physical magnitudes. These quantum moments are also useful to implement a consistent notion of hybrid entropy in a simple way, since they take the form of a family of density matrices indexed by the classical degrees of freedom, which encode in a simple way the exclusivity of quantum (and hence hybrid) events.  In future works, we plan to analyze in detail the implications of these new entropy functions, and some relevant examples of ensembles arising from them, such as the canonical or micro-canonical ensembles. Furthermore, the stability of the resulting distributions with respect to the dynamics can now be studied in a simple way. Following that direction, the analysis of equilibrium thermodynamics of hybrid systems is a feasible objective. 

We have also been able to re-write the Liouville Ehrenfest dynamics of the full probability density at the level of the quantum moments. We have obtained a series of coupled differential equations for the set of quantum moments to encode the physical content of Liouville equation. Furthermore, using the fact that only a finite number of quantum moments are required to model the system for finite time intervals, we have considered the problem of defining the dynamics with a finite number of those equations. In order to do that, we need to write the effect of the next moments on the highest one. As those higher moments  can not be determined by physical measurements, we choose to represent those  degrees of freedom by the state, compatible with the constraints of the problem, which maximizes the hybrid entropy at that level.  With this choice, a well defined differential equation can be written for the system.

An important consequence of our results is the change it implies at the level of implementing numerical simulations of statistical hybrid systems. We have learned that an accurate description of a statistical hybrid system requires taking into account the correlation between quantum observables mediated by the nonlinear classical subsystem, i.e., the enlargement of the algebra of hybrid observables. This property incorporates into the description more quantum moments which are not required to obtain the initial average values of physical magnitudes, but that are necessary to estimate their evolution. From the point of view of numerical simulation this introduces some difficulties, since incorporating these higher order correlations into the usual simulation methods as the use of independent (pure-state) trajectories is not an easy task. While fixing values of single operators is simple, considering correlations of several magnitudes represents a new numerical challenge.
The issue of knowing the statistical value of the correlations between observables is closely related to the issue of preparation of statistical quantum (in this case, hybrid) states. 
On the other hand, the preparation of states is very often implemented in simulations through a sampling over quantum phase space of initial conditions that reproduce the average initial values of the characterized observables. Our results show that for non linear dynamics it is not feasible to just approximate the quantum state as a bundle of trajectories with a statistical sampling over quantum phase space of initial conditions, knowing just the expectation value of operators. The main difference with  von Neumann's evolution is that such master equation given in terms of $\hat\rho$ is linear on quantum projectors and such linearity preserves the original sampling throughout the evolution.  In this hybrid case, such sampling must reproduce correlations and higher order observables up to the desired level of precision (associated with $k+1$ in the truncation above), or work directly to the set of $k$ density matrices, characterized through such measurements, to be able to implement the coupled evolution between subsequent powers of quantum projectors in such non-linear dynamics.


\section*{Acknowledgements}
The authors acknowledge partial finantial support of  Grant PID2021-123251NB-I00 funded by MCIN/AEI/10.13039/501100011033  and by the European Union, and of Grant E48-23R funded by Government of Aragón. C.B-M and D.M-C acknowledge financial support by Gobierno de Aragón through the
grants defined in ORDEN IIU/1408/2018 and ORDEN CUS/581/2020 respectively.

\section*{Data Availability Statement}
No Data associated in the manuscript

\end{document}